\documentclass[journal]{IEEEtran}

\usepackage[table]{xcolor}      
\usepackage{array}              
\usepackage[pdftex]{graphicx}   
\usepackage{url}                
\DeclareGraphicsExtensions{.pdf,.jpeg,.png} 
\usepackage[cmex10]{amsmath}    
\usepackage{mdwmath}            
\usepackage{mdwtab}             
\usepackage{eqparbox}           
\usepackage{amsmath}            
\usepackage{multirow}           
\usepackage{booktabs}           
\usepackage{adjustbox}          
\usepackage{multicol}           
\newcolumntype{a}{>{\columncolor{Gray}}c}
 \usepackage{threeparttable} 

\begin{document}
\bstctlcite{IEEEexample:BSTcontrol}
    \title{System-Level Analysis for mm-Wave Full-Duplex Transceivers }
  \author{ Mohamad Mahdi~Rajaei Rizi,~\IEEEmembership{Graduate Student Member,~IEEE,}            Jeyanandh~Paramesh,~\IEEEmembership{Senior Member,~IEEE,}
      and Kamran~Entesari,~\IEEEmembership{Senior Member,~IEEE} 

\thanks{The authors are with the Department of Electrical and Computer Engineering, Texas A\&M University, College Station, TX, 77843 USA (e-mail: m.rajaei94@tamu.edu).}
\thanks{This work was supported by the National Science Foundation (NSF) under Award 2116498.}}


\maketitle

\begin{abstract}
This paper conducts a comprehensive system-level analysis of mm-Wave full-duplex transceivers, focusing on a receiver employing a four-stage self-interference cancellation (SIC) process. The analysis aims to optimize the noise and linearity performance requirements of each transceiver block, ensuring that the self-interference (SI) signal does not compromise the receiver's error vector magnitude (EVM) for an OFDM 64-QAM signal. Additionally, the necessary SIC for each stage is calculated to establish feasible noise and linearity specifications for a CMOS-based implementation. The resulting specifications are subsequently validated within a MATLAB Simulink environment, confirming the accuracy of the computed requirements for each block.
\end{abstract}


\IEEEpeerreviewmaketitle


\section{Introduction}\label{sec:num1}
\IEEEPARstart{F}{ull-Duplex} wireless systems enable simultaneous transmission and reception over a single antenna, improving system compactness and offering substantial benefits in spectral efficiency, link capacity, and network latency reduction. By allowing both the transmitter (TX) and receiver (RX) to operate concurrently on the same frequency, these systems—referred to as in-band full-duplex (IBFD)—excel in supporting high data rates, particularly in densely populated network environments \cite{FD_intro_1}. This concurrent operation provides significant advantages over traditional duplexing methods like frequency-division duplexing (FDD) and time-division duplexing (TDD). Beyond physical-layer enhancements, full-duplex technology delivers additional performance gains across higher network layers, as discussed in references \cite{FD_intro_1}, \cite{FD_intro_2}, \cite{FD_intro_3}, \cite{FD_intro_4}, \cite{FD_intro_5}, and \cite{FD_intro_6}. However, a major design challenge is the suppression of the self-interference (SI) signal, which can be orders of magnitude stronger than the desired signal—potentially up to a million times—posing risks of receiver desensitization and severe signal-to-noise ratio (SNR) degradation. To address this, a single cancellation stage is typically insufficient; instead, a multi-stage self-interference cancellation approach is required, each stage providing targeted SI attenuation to meet system requirements. A detailed link-budget analysis is therefore essential, enabling the design of receiver components with realistic noise and linearity specifications that ensure the SI level is minimized to achieve the desired SNR. Furthermore, achieving high noise and linearity performance is challenging in CMOS technologies at mm-wave frequencies, necessitating meticulous optimizations to extract sensible values for each SIC stage and each block's specifications. In this regard, this work presents a detailed system-level analysis of a mm-wave FD transceiver verified by modulation-based simulations in Simulink environment.

\section{Proposed System-Level Analysis}\label{sec:num2}

\begin{figure}[t]
  \begin{center}
  \includegraphics[width=3.5in]{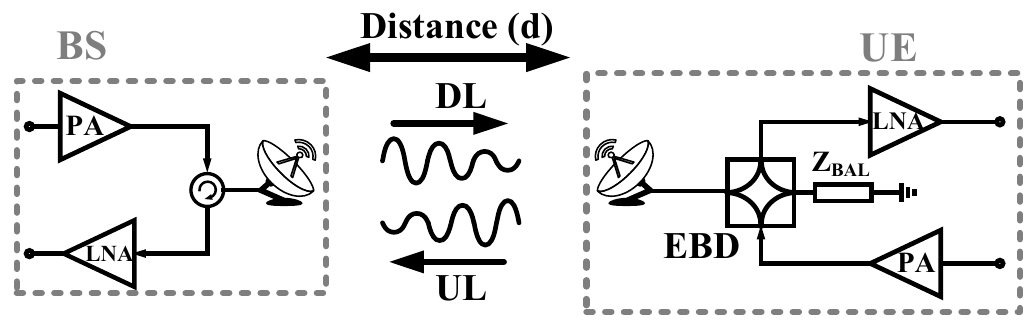}\\
  \caption{A full-duplex communication link.}\label{System_1}
  \end{center}
\end{figure}

Fig.~\ref{System_1} illustrates a full duplex link implementing the communication between a base station (BS) and a user equipment (UE) front-ends. The separation distance between the BS and UE is denoted by 'd'. Additionally, the transceiver at the UE end comprises a power amplifier and a low noise amplifier connected to an EBD aiming to isolate transmission\,(TX) and reception\,(RX) paths. However, the isolation provided by the EBD is limited, leading to the presence of self-interference\,(SI) at the receiver input. In full-duplex\,(FD) systems, where both transmitter and receiver operate concurrently at the same frequency, it is essential to adequately suppress the transmitter leakage signal across multiple stages of self-interference cancellation (SIC) to achieve targeted EVM. As a result, meticulous budget analysis becomes vital to determine optimal values for noise and linearity metrics of each block, alongside the SIC value of each cancellation node across the receiver chain. 

The analysis will be conducted based on two scenarios: uplink and downlink. In the uplink scenario, the transmission from the UE and reception at the BS are addressed, whereas in the downlink scenario, the situation is reversed. Note that in both scenarios the existence of SI affects the reception. The focus of this article, however, is the reception performance of the UE side. Hence, SI influences and mitigation methods are addressed only at downlink scenario. Each scenario is further elaborated upon in the subsequent subsections.  For a comprehensive understanding of the parameters utilized throughout the analysis, readers can refer to Table I.
\begin{table}
    \centering
    \caption{System Parameters}
    \begin{tabular}{|c|c|}
        \hline
        Parameter & Description \\
        \hline
        \hline
        $P_{\text{TX,BS(UE)}}$ & Power delivered to the antenna at the BS\,(UE) side\\
        \hline
        $P_{\text{RX,BS(UE)}}$ & Power collected by the antenna at the BS\,(UE) side\\
        \hline        
        $G_{\text{BS(UE)}}$ & BS\,(UE) antenna gain \\       
        \hline     
        $L_{\text{FS}}$ & Free space path loss\\
        \hline
        $NF_{\text{BS(UE)}}$ & NF of the (UE)\,BS's receiver\\
        \hline
        $BW_{\text{DL(UL)}}$ & DL\,(UL) Modulation band-width\\
        \hline
        $SNR_{\text{RX}}$ & Minimum output SNR requirement for a given EVM\\
        \hline        
    \end{tabular}    
    \label{SYS_parameter_def}
\end{table}
\subsection{Uplink}
In this scenario, the required output power transmitted by the UE will be determined. The modulation scheme considered for the uplink is an OFDM 64-QAM with 400\,MHz bandwidth necessitating a total link SNR of 21\,dB\,\cite{SNR_64QAM}. To allocate this SNR value between the transmitter\,(UE) and the receiver\,(BS), (\ref{SNR_link}) is utilized \cite{SNR_64QAM}:
 \begin{equation}
\begin{split}
    SNR_{\text{Link}} &= \frac{1}{\frac{1}{SNR_{\text{TX}}}+\frac{1}{SNR_{\text{RX}}}}                      
    \label{SNR_link}
    \end{split}
\end{equation}
The $\text{SNR}_{\text{TX}}$ is primarily attributed to PA non-linearity effects, with a typical value based on the reported mm-wave PAs in literature being 24\,dB \cite{PA_1,PA_2,PA_3}. Thus, employing (\ref{SNR_link}):

 \begin{equation}
\begin{split}
    SNR_{\text{RX}} &= 24\,\text{dB}                      
    \label{SNR_rx}
    \end{split}
\end{equation}

This value for $\text{SNR}_{\text{RX}}$ represents the ratio of signal to all non-ideal effects (such as noise and non-linearity). Considering that the undesired signal components produced by other non-ideal effects impose a 3\,dB margin on the link budget, then the $\text{SNR}_{\text{noise}}$ mentioned in (\ref{P_min_bs}) will be 27\,dB. 
\begin{equation}
\begin{split}
    P_{\text{RX,BS,min}} =&-174\,dBm/Hz + 10 \log\left(BW_{\text{UL}}\right)\\
                          & + NF_{\text{BS}}+ SNR_{\text{noise}} 
    \label{P_min_bs}
    \end{split}
\end{equation}
Given a NF of 8\,dB for the BS, a  minimum received power of -53\,dBm is necessary.

The relationship between the received power at the BS and the transmitted power by the UE is expressed by the equation below:
\begin{equation}
\begin{split}
    P_{\text{RX,BS}} &= P_{\text{TX,UE}} + G_{\text{UE}} - L_{\text{FS}} + G_{\text{BS}}
\label{uplink}
\end{split}
\end{equation}
Using (\ref{uplink}) and assuming a gain of 20\,dBi\,(=\,$\text{G}_{\text{BS}}$\,=\,$\text{G}_{\text{UE}}$) for an aperture mm-wave antenna\,\cite{Antenna_1} with a maximum distance of 90\,m between the BS and the UE, leading to an $L_{\text{FS}}$ of 101\,dB, we obtain:
\begin{equation}
\begin{split}
    P\,_{\text{TX,UE}} & >  8\,\text{dBm}                  
    \label{PTXUE}
    \end{split}
\end{equation}
With a typical insertion loss of 4\,dB for EBD TX path, the output power of the PA at the UE side must be: 
\begin{equation}
\begin{split}
    P\,_{\text{PA-out,UE}} & >  12\,\text{dBm}                  
    \label{PPAoutUE}
    \end{split}
\end{equation}
\subsection{Downlink}
\begin{figure}[t]
  \begin{center}
  \includegraphics[width=3.5in]{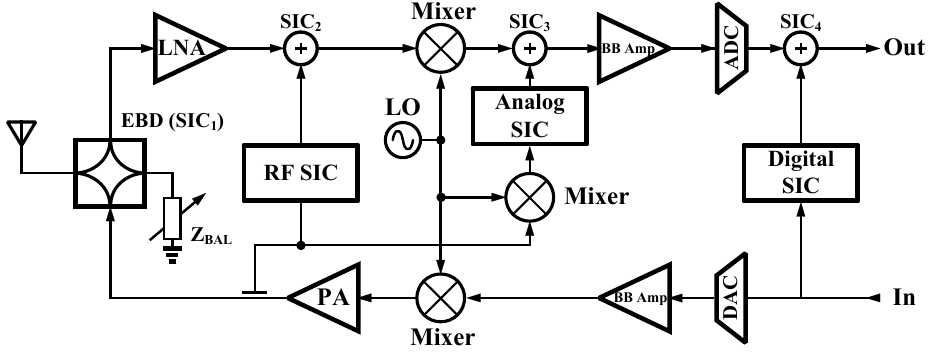}\\
  \caption{Full-duplex transceiver block-diagram at UE end.}\label{System_2}
  \end{center}
\end{figure}

In this scenario, the analysis focuses on the UE reception performance in the presence of the SI. Typical noise and linearity performances for mm-wave CMOS receivers will be considered to ensure that the EVM (SNR) requirement for the assumed modulation scheme is fulfilled. As depicted in Fig.~\ref{System_2}, the UE receiver incorporates various stages of SI-cancellation, each offering specific advantages to the entire system. Determining the appropriate value requires thorough considerations, and different criteria in the following sub-sections  are introduced to extract $\text{SIC}_{\text{i}}$ values.

\subsubsection{Total Required SIC}\label{sssec:num1}
The criterion here is to find the $\text{SIC}_{\text{total}}$ such that the power of residual SI signal after all SIC stages is adequately below the noise level. The Modulation scheme for downlink scenario is assumed to be an OFDM 64-QAM with 400\,MHz band-width. Thus, the minimum required $SNR_{\text{Link}}$ is 21\,dB\,\cite{SNR_64QAM}. The definition of SNR in this context is identical with that of the uplink scenario. Similarly, to distribute this total SNR value between the BS and UE, equation (\ref{SNR_link}) can be utilized. Note that in this system-level analysis, we assume that the PA at the UE side provides lower linearity performance compared to the BS side due to general design limitations that are subjected to UE side. Hence, the $SNR_{\text{TX(BS)}}$ is assumed to be 30\,dB based on PA specs mentioned in \cite{PA_3} leading to an $SNR_{\text{RX(UE)}}$ of 21.6\,dB.

To apportion the $SNR_{\text{RX(UE)}}$ among noise, non-linearity of the receiver, and residual SI, equation (\ref{SNR_RX}) is employed\,\cite{SNR_64QAM}. In this equation, the effect of IM3 components produced by mixer and other subsequent stages are ignored; the dominance of the IM3 components generated by the LNA stems from the design of $\text{SIC}_{\text{2}}$ as discussed in section \ref{sssec:num3}, which will be tailored to mitigate the IM3 components of subsequent stages (Mixer and base-band amplifier). Therefore:
\begin{equation}
\begin{split}
    SNR_{\text{RX(UE)}} &= \frac{1}{\frac{1}{SNR_{\text{noise}}}+\frac{1}{SNR_{\text{IM3-LNA}}}+\frac{1}{SNR_{SI}}}                      
    \label{SNR_RX}
    \end{split}
\end{equation}
\\
Where $SNR_{\text{noise}}$ represents the common signal-to-noise ratio, $SNR_{\text{IM3-LNA}}$ denotes the ratio of the desired signal to IM3 components produced by the LNA, and the $SNR_{\text{SI}}$ signifies the signal to the input referred residual SI level ratio, as defined in (\ref{SNR_SI_def}).
\begin{equation}
\begin{split}
    SNR_{\text{SI}}\,&=P_{\text{RX,UE}}-P_{\text{SI,residual}} 
    \label{SNR_SI_def}
    \end{split}
\end{equation}
Where $\text{P}_{\text{RX,UE}}$ is determined in (\ref{eq1})
\begin{equation}
\begin{split}
    P_{\text{RX,UE}} &= P_{\text{TX,BS}} + G_{\text{BS}} - L_{\text{FS}} + G_{\text{UE}}                     
    \label{eq1}
    \end{split}
\end{equation}
In previous studies \cite{PA_1}, \cite{PA_2}, and \cite{PA_3}, CMOS mm-wave PAs supporting modulation schemes such as 64-QAM and/or 256-QAM with output powers reaching up to 15\,dBm are presented. Hence, 15\,dBm is a reasonable choice for $P_{\text{TX,BS}}$; thus, the minimum received power calculated by (\ref{eq1}) is:
\begin{equation}
\begin{split}
    P_{\text{RX,UE}} &= -46\,\text{dBm}                      
    \label{eq2}
    \end{split}
\end{equation}
Note that a general purpose receiver might indicate better $\text{P}_{\text{RX,UE}}$ than the value achieved in (\ref{eq2}), but in the context full-duplex receiver obtaining very sensitive receivers requires super-linear LNA and prohibitively high value of $\text{SIC}_{\text{1}}$ [see Fig.~\ref{System_2}] which is very challenging at mm-wave frequencies. However, the methodology and fellow of analysis proposed here is valid while values can change according to technology of fabrication and frequency of operation.

Equation (\ref{P_min_bs}) can be re-formulated for the downlink, yielding an 8\,dB NF for UE with an $\text{SNR}_{\text{noise}}$ of 28\,dB. The "sufficient" value of SI cancellation will be achieved when:
\begin{equation}
\begin{split}
    \frac{1}{SNR_{SI}}<<(\frac{1}{SNR_{\text{noise}}}+\frac{1}{SNR_{\text{IM3-LNA}}})         
    \label{SNR_SI_cri_0}
    \end{split}
\end{equation}
To validate this approximation, we assume:
\begin{equation}
\begin{split}
    \frac{1}{SNR_{SI}}= (\frac{1}{SNR_{\text{noise}}}+\frac{1}{SNR_{\text{IM3-LNA}}})\times0.01         
    \label{SNR_SI_cri}
    \end{split}
\end{equation}
By neglecting $\text{SNR}_{\text{SI}}$ in (\ref{SNR_RX}), this equation necessitates an $SNR_{\text{IM3-LNA}}$ of 23\,dB to meet the 21.6\,dB requirement for the entire RX. Thus, by substitution $SNR_{\text{noise}}$ and $SNR_{\text{IM3-LNA}}$ into (\ref{SNR_SI_cri}), $SNR_{\text{SI}}$ is determined as follows:
\begin{equation}
\begin{split}
    SNR_{\text{SI}}\,&=44\,\text{dB}
    \label{SNR_SI_num}
    \end{split}
\end{equation}
By definition, the total required SIC is:

\begin{equation}
\begin{split}
    SIC_{\text{total}}\,&=\,P\,_{\text{PA-out,UE}}-\underbrace {P\,_{\text{SI-residual}}}_{(P_{\text{RX,UE}}-SNR_{\text{SI}})\,\text{(\ref{SNR_SI_def})}} 
    \label{sic_t}
    \end{split}
\end{equation}

Using Eqs.(\ref{PPAoutUE}, \ref{eq2}, \ref{SNR_SI_num}, and\ref{sic_t}) the total required SIC is:
\begin{equation}
\begin{split}
    SIC_{\text{total}}\,&>\,102\,\text{dB}
    \label{sic_t_num}
    \end{split}
\end{equation}
\subsubsection{$\text{SIC}_{\text{1}}$ (EBD Isolation)}\label{sssec:num2}
The criterion for a well designed $\text{SIC}_{\text{1}}$ is to sufficiently reduce the SI signal to a level where the IM3 components of the LNA, with given noise and linearity specifications, do not degrade $\text{SNR}_{\text{IM3-LNA}}$ from 23\,dB. 

The $\text{SNR}_{\text{IM3-LNA}}$ is calculated as below:
\begin{equation}
\begin{split}
    SNR_{\text{IM3-LNA}} &= P_{\text{RX,UE}} - P_{\text{IM3-LNA}}                      
    \label{SNR_IM3}
    \end{split}
\end{equation}
In which $P_{\text{RX,UE}}$ is determined in (\ref{eq2}), and $P_{\text{IM3-LNA}}$ is the input-referred power of IM3 components generated by the LNA due to the strongest signal applied to the receiver, which, in this case, is the SI signal after first SIC stage and will be denoted by $P_{\text{SI-1}}$. 

\begin{figure}[t]
  \begin{center}
  \includegraphics[width=3.25in]{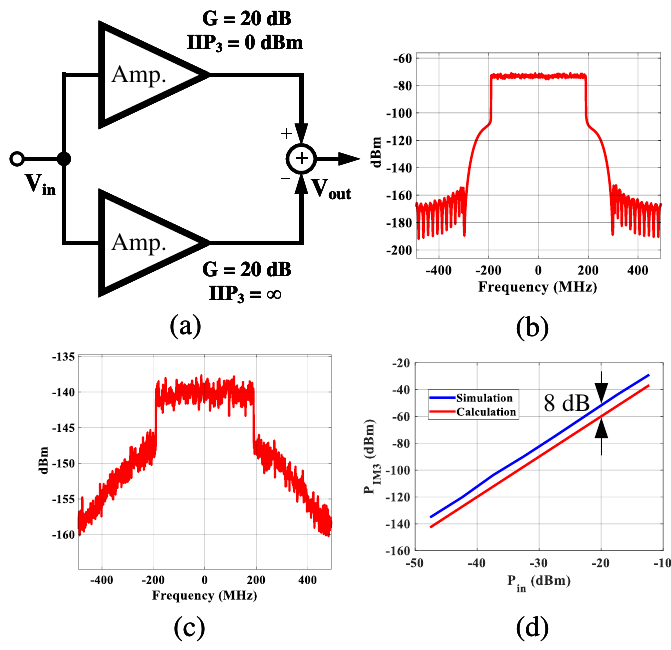}\\
  \caption{(a) Simulation setup implemented in Simulink to extract the correction factor (b) Input spectrum with 400\,MHz and -47.4\,dBm occupied bandwidth and average channel power, respectively. (c) Output spectrum of the setup signifying a channel power level of -114.7\,dBm for the IM3 products. (d) Input-referred power level of IM3 products versus input power, the blue curve represents the simulation results and red curve is the predicted value by (\ref{iip3_eq}).}\label{SYS_IM3}
  \end{center}
\end{figure}

$P_{\text{IM3-LNA}}$ can be related to the $\text{IIP}_{\text{3}}$ of the LNA as below, provided that the SI is modeled with two continuous wave (CW) signals:
\begin{equation}
\begin{split}
    P_{\text{IM3-LNA}} &= 3P_{\text{SI-1}}-2IIP_{\text{3-LNA}}
    \label{iip3_eq}
    \end{split}
\end{equation}
However, the assumption based on which (\ref{iip3_eq}) is extracted is too simplistic for this modulation-based budget analysis as the instantaneous power time variation are considerable and substitution of average power in (\ref{iip3_eq}) underestimates the power level of IM3 products. Intuitively, the power level of the IM3 products can be confined between the below limits:

{\small
\begin{equation}
\begin{split}
    &3P_{\text{SI-1}}-2IIP_{\text{3-LNA}}< P_{\text{IM3-LNA}}<3(P_{\text{SI-1}}+PAPR)-2IIP_{\text{3-LNA}}
    \label{iip3_eq_1}
    \end{split}
\end{equation}
}
In which PAPR determines the peak to average power ratio and is a function of the modulation scheme. To estimate the IM3 power with acceptable accuracy a correction should be applied to (\ref{iip3_eq}). In this regard, MATLAB envelope simulation for an OFDM 64-QAM 400\,MHz signal with a sub-carrier spacing (SCS) of 120\,kHz, carried out in Simulink environment as displayed in Fig.~\ref{SYS_IM3}(a); in which two amplifiers with the identical gain values are excited with the similar output, and the subtracted output are measured to extract the IM3 power. Figs.~\ref{SYS_IM3}(b and c) illustrate the spectrum of the input and output signals of the setup shown in Fig.~\ref{SYS_IM3}(a), respectively. Fig.~\ref{SYS_IM3}(d) shows the input-referred power of the IM3 products versus input power for estimated by simulation (blue curve) and (\ref{iip3_eq}), resulting in an 8\,dB correction. Thus, we arrive at:  
\begin{equation}
\begin{split}
   P_{\text{IM3-LNA}} &= 3P_{\text{SI-1}}-2IIP_{\text{3-LNA}}+8
    \label{iip3_eq_mod}
    \end{split}
\end{equation}

By substituting (\ref{iip3_eq_mod}) into (\ref{SNR_IM3}), the $P_{\text{SI-1}}$ is expressed as below:
\begin{equation}
\begin{split}
    P_{\text{SI-1}} < \frac{1}{3}\times(P_{\text{RX,UE}}-SNR_{\text{IM3-LNA}}+2IIP_{\text{3-LNA}}-8)
\label{PSI_1}
\end{split}
\end{equation}

By definition, $\text{SIC}_\text{1}$ is:
\begin{equation}
\begin{split}
    SIC_1 &> P\,_{\text{PA-out,UE}} - \frac{1}{3}\times(P_{\text{RX,UE}}-SNR_{\text{IM3-LNA}}\\
        &+2IIP_{\text{3-LNA}}-8)
\label{SIC-1}
\end{split}
\end{equation}

Assuming an $IIP_{3}$ of -7\,dBm for the LNA and using (\ref{PPAoutUE} and \ref{eq2}), to obtain an $\text{SNR}_{\text{IM3-LNA}}$ of 23\,dB, the $SIC_1$ should be:
\begin{equation}
\begin{split}
    SIC_1 &> 42\,\text{dB}
\label{SIC_1_formula}
\end{split}
\end{equation}

\subsubsection{$\text{SIC}_{\text{2}}$ (RF SIC)}\label{sssec:num3}
The criterion to calculate $\text{SIC}_{\text{2}}$ is that the SI level must be sufficiently low to ensure that the IM3 components generated by mixer and base-band (BB) amplifier are significantly below IM3 components produced by the LNA. To meet this requirement and to cover the underestimation issue mentioned in section \ref{sssec:num2}, a 18\,dB margin is taken into account. This margin accounts for the 8\,dB underestimation of IM3 power, plus an additional 10\,dB to ensure that the aforementioned criterion is adequately satisfied. Consequently:
\begin{equation}
\begin{split}
    P\,_{\text{IM3-RRX}} < P\,_{\text{IM3-LNA}}\,+\,G_{\text{LNA}}\,-\,18\,\text{dB}
\label{Im3_Criterion}
\end{split}
\end{equation}
In which $P\,_{\text{IM3-RRX}}$ denotes the IM3 components generated by the rest of the RX (RRX) and is determined as follows:
\begin{equation}
\begin{split}
    P_{\text{IM3-RRX}} &= 3(P_{\text{SI-1}}\,+\,G_{\text{LNA}}\,-\,SIC_{\text{2}})\,-\,2IIP_{\text{3-RRX}}
    \label{P_IM3_rrx}
    \end{split}
\end{equation}
Substituting (\ref{P_IM3_rrx}) into (\ref{Im3_Criterion}), we arrive at:
\begin{equation}
\begin{split}
    SIC_{\text{2}} &> \frac{2}{3}\times(OIP_{\text{3-LNA}}\,-\,IIP_{\text{3-RRX}})+6
    \label{sic2}
    \end{split}
\end{equation}

Assuming a gain of 20\,dB for LNA and an $\text{IIP}_{\text{3-RRX}}$ of -15\,dBm, 25\,dB of SIC is adequate for the second stage of cancellation.
\subsubsection{$\text{SIC}_{\text{3}}$ (Analog SIC)}
The role of this stage of the SI-canceller is twofold: First, it addresses the issue of IM3 components from the PA that leak into the RX, contributing to SNR degradation, along with the primary SI signal. Given the unavailability of access to these IM3 components in the digital domain, this stage is crucial to ensure adequate reduction of the PA-generated IM3 components. Second, being the final stage of analog domain SIC, it is essential to ensure that the residual power of the SI signal falls within the dynamic range of the ADC.

For PA-generated IM3 components, a 10\,dB margin is assumed, thus:
\begin{equation}
\begin{split}
    &P_{\text{OIM3-PA}}-SIC_1+G_{\text{LNA}}-SIC_2+G_{\text{Mixer}}+G_{\text{BBAmp}}\\
    &-SIC_3<P_{\text{IM3-LNA+Noise}}+G_{\text{LNA}}+G_{\text{Mixer}}+G_{\text{BBAmp}}+10  
\label{sic3_1}
\end{split}
\end{equation}
 Where $P_{\text{OIM3-PA}}$ is the IM3 at the output of the PA, and $P_{\text{IM3-LNA+Noise}}$ signifies the collective power of input-referred receiver noise and input-referred LNA IM3 products in dBm. Simplifying (\ref{sic3_1}), the required $\text{SIC}_\text{3}$ is calculated as follows:
 \begin{equation}
\begin{split}
    &SIC_3>P_{\text{OIM3-PA}}-P_{\text{IM3-LNA+Noise}}-SIC_1-SIC_2+10
\label{sic3_2}
\end{split}
\end{equation}
In essence, (\ref{sic3_2}) is general, and $P_{\text{OIM3-PA}}$ encompasses any undesired components produces by power amplifier, without the restriction of being strictly third-order non-linearity. 

Leveraging the PA specifications that will be presented in this article with an $\text{OP}_{\text{1dB}}$ of +15\,dBm with a power gain of 13.5\,dB, a high-level MATLAB Simulink model is produced with the setup shown in Fig.~\ref{SYS_IM3}(a) yielding a simulated $P_{\text{OIM3-PA}}$ of 0\,dBm. Using (\ref{sic3_2}), the required $\text{SIC}_\text{3}$ is 16\,dB. 

To verify the second criterion mentioned above, the remaining SIC (i.e. digital SIC) must be less than ADC dynamic range. Thus, 
\begin{equation}
\begin{split}
      DR_{\text{ADC}}=6\times(ENOB-2)>(SIC_{total}-\Sigma SIC_i)
\label{DR_ADC}
\end{split}
\end{equation}

Where ENOB is the ADC effective number of bits and assumed to be 8\,bit. Hence, by substitution of above-calculated values into (\ref{DR_ADC}), it is assured that the $\text{SIC}_{\text{3}}$ calculated based on the first criterion fulfills the second criterion as well.  

\subsubsection{$\text{SIC}_{\text{4}}$ (Digital SIC)}
The digital SIC can be calculated by (\ref{sic4_1}), resulting in 17\,dB digital SIC requirement.  
\begin{equation}
\begin{split}
    &SIC_4>SIC_{\text{total}}-(SIC_1+SIC_2+SIC_3)
\label{sic4_1}
\end{split}
\end{equation}

\begin{figure}[t]
  \begin{center}
  \includegraphics[width=3.5in]{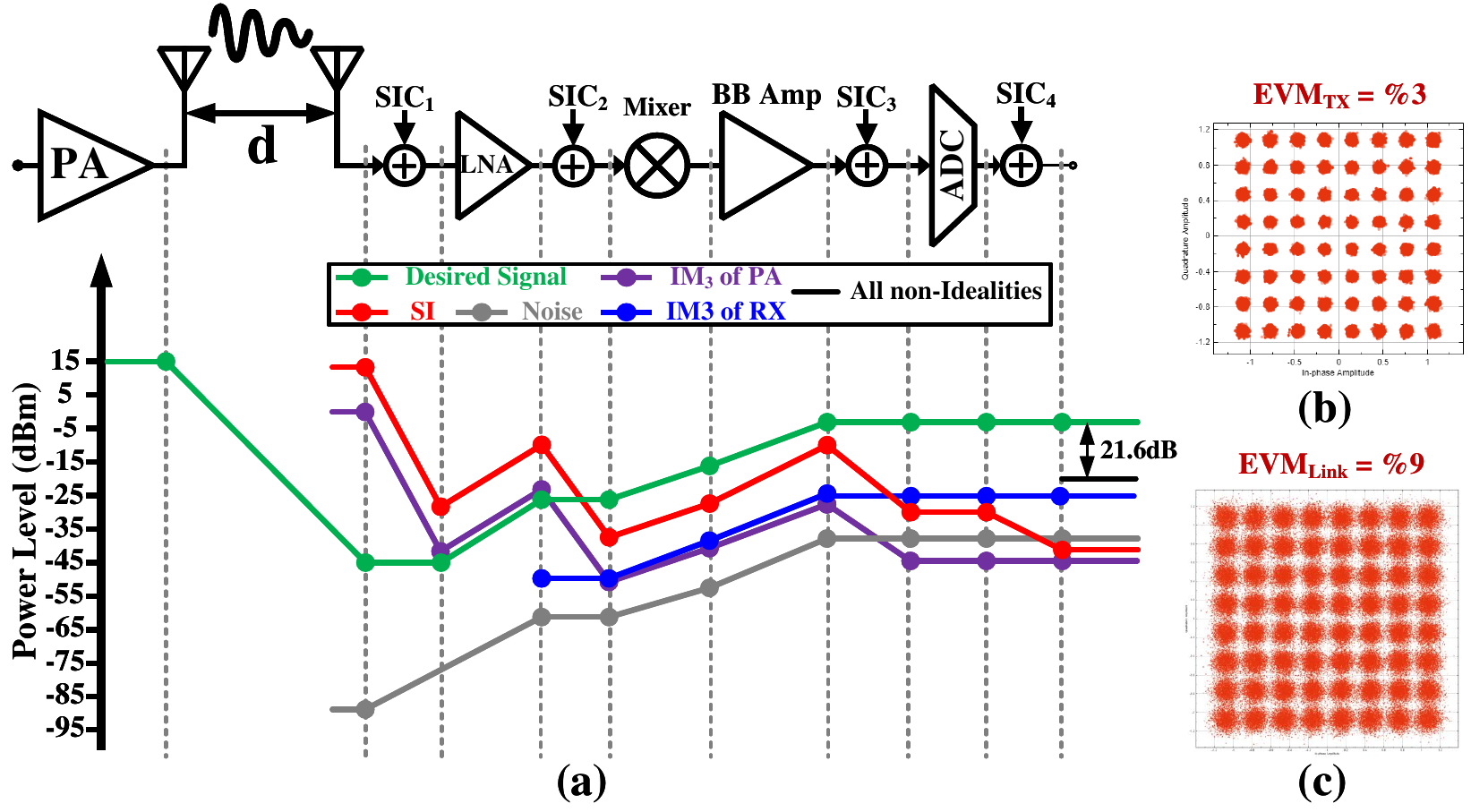}\\
  \caption{(a) The track of simulated power levels for desired signal, SI, IM3 product generated by RX, IM3 product generated by UE's PA, and noise through the link chain ($\text{SIC}_{\text{1}}=\text{40\,dB}$,  $\text{SIC}_{\text{2}}=\text{28\,dB}$, $\text{SIC}_{\text{3}}=\text{16\,dB}$, $\text{SIC}_{\text{4}}=\text{10\,dB}$, $\text{NF}_{\text{RX}}=\text{8\,dB}$, $\text{IIP}_{\text{3-LNA}}=\text{-7\,dBm}$, $\text{IIP}_{\text{3-Mixer}}=\text{0\,dBm}$, $\text{IIP}_{\text{3-BBAmp}}=\text{+5\,dBm}$, and $\text{OP}_{\text{1dB-PA(UE)}}=\text{+15\,dBm}$). (b) Simulated constellation of the signal from the BS end; the resulted $\%\text{EVM}_{\text{rms}}$ is $\%$3. (c) Simulated constellation of the entire link; the resulted $\%\text{EVM}_{\text{rms}}$ is $\%$9.
  \label{System_3}}
  \end{center}
\end{figure}

A comprehensive model of the link with the aforementioned  assumptions for each block specification is produced in MATLAB Simulink environment. Fig.~\ref{System_3}(a) illustrates the simulated power level at different nodes through the link. The PA output signal representing the BS transmitter with high-level specs reported in \cite{PA_3} indicates simulated $\text{EVM}_\text{rms}$ of $\%$3, which is consistent with the results mentioned in \cite{PA_3}. Fig.~\ref{System_3}(b) shows the signal constellation at the output of the TX side of the link, verifying the assumption of $SNR_{\text{TX(BS)}}$ = 30\,dB discussed in section \ref{sssec:num1}. Moreover, the signal constellation of the entire link is depicted in Fig.~\ref{System_3}(c), and the resulted $\text{EVM}_\text{rms}$ for the whole link is $\%$9  reinforcing the entire presented link analysis.   

This article aims to propose a solution for $\text{SIC}_{\text{1}}$ using electrical balance duplexer integrating a PA and LNA in 22\,nm CMOS FDSOI which provides enough transmitted power, TX-RX isolation, and RX path gain across the bandwidth. The following sections will elaborate design, analysis, and measurement of the proposed duplexer.

\section{Conclusion}\label{sec:num7}
This work presents an in-depth system-level investigation into mm-Wave full-duplex transceivers, particularly analyzing a receiver that integrates a four-stage self-interference cancellation (SIC) mechanism. The study targets the optimization of noise and linearity performance across each component of the transceiver to prevent the self-interference (SI) signal from impacting the error vector magnitude (EVM) for a 64-QAM modulation scheme. Furthermore, each SIC stage's required cancellation level is determined to support practical noise and linearity targets suitable for CMOS technology. These calculated specifications are then verified through simulations in MATLAB Simulink, confirming that each transceiver block meets the outlined design criteria.
\ifCLASSOPTIONcaptionsoff
  \newpage
\fi
\bibliographystyle{IEEEtran}
\bibliography{Bibliography}
\vfill
\end{document}